\documentclass[12pt]{article}

\usepackage{graphicx}
\usepackage{amsmath,amsthm,amssymb}
\usepackage{maple2e}
\usepackage{pb-diagram}
\chardef\bslash=`\\ 

\theoremstyle{definition}

\newtheorem{conj}{Conjecture}
\theoremstyle{remark}

\newcommand{\eval}[2][\right]{\relax
 \ifx#1\right\relax \left.\fi#2#1\rvert}

\begin{document}
\title{\sf{Formation of singularities in
Yang-Mills equations}}
\author{Piotr Bizo\'n\\
 \small{\textit{Institute of Physics,
Jagellonian University, Krak\'ow, Poland}}}

\pagestyle{headings}

\maketitle
\begin{abstract}
\noindent This is a survey of recent studies  of singularity
formation in solutions of spherically symmetric Yang-Mills
equations in higher dimensions.  The main attention is focused on
five space dimensions because this case exhibits interesting
similarities with Einstein's equations in the physical dimension,
in particular the dynamics at the threshold of singularity
formation shares many features (such as universality,
self-similarity, and scaling) with critical phenomena in
gravitational collapse.  The borderline case of four space
dimensions is also analyzed and the formation of singularities is
shown to be intimately tied to the existence of the instanton
solution.

\end{abstract}
\section{Introduction}
One of the most interesting features of many nonlinear evolution
equations is the spontaneous onset of singularities in solutions
starting from perfectly smooth initial data. Such a phenomenon,
usually called "blowup", has been a subject of intensive studies
in many fields ranging from fluid dynamics to general relativity.
Whether or not the blowup can occur for a given nonlinear
evolution equation is the central mathematical question which,
from the physical point of view, has a direct bearing  on our
understanding of the limits of validity of the corresponding
model. Unfortunately, this is often a difficult question. Two
famous examples for which the answer is not known are the
Navier-Stokes equation and the Einstein equations. Once the
existence of blowup is established for a particular equation, many
further questions come up, such as: When and where does the blowup
occur? What is the character of blowup and is it universal? Can a
solution be continued past the singularity?

In this paper we consider these questions for the Yang-Mills (YM)
equations in higher dimensions. In the physical $3+1$ dimensions,
where the YM equations are the basic equations of gauge
 theories  describing the weak and strong
 interactions of elementary particles, it is known
 that no singularities can form.
 This was shown
by Eardley and Moncrief~\cite{ea_mo} who proved that solutions
starting from smooth initial data remain smooth for all future
times. The motivation for studying the YM equations in unphysical
$D+1$ dimensions for $D>3$ is twofold (and unrelated to the latest
fashion of doing physics in extra dimensions). From the
mathematical point of view, it is the obvious thing to ask
  how
the property of global regularity depends on the dimension of the
underlying spacetime and whether  singularities can form in $D+1$
dimensions for $D>3$.
However, there is also  a  less evident physical reason which is
motivated by the hope
 that by understanding the problem of
singularity formation for the YM equations one might get insight
into the analogous, but much more difficult, problem in general
relativity. From this viewpoint -- in which the YM equations are
considered  as a toy model for the Einstein equations -- it is
essential that these two equations  belong to the same criticality
class. Let us recall that the criticality class is defined as the
degree $\alpha$ in the homogeneous scaling of energy $E
\rightarrow \lambda^{\alpha} E$ under dilations  $x \rightarrow
x/\lambda$. The classification of equations into subcritical
($\alpha<0$), critical ($\alpha=0$), and supercritical
($\alpha>0$) is a basis of  the heuristic meta-principle according
to which subcritical equations are globally regular, while
supercritical equations may develop singularities for some (large)
initial data~\cite{kl1}. For the YM equations we have
$\alpha_{YM}=D-4$, while for the Einstein equations
$\alpha_{E}=D-2$. Therefore, the YM equations in $D=5$ have the
same criticality, $\alpha=1$, as the Einstein equations in the
physical dimension.  Another way of seeing this is to note that in
$D=5$ the dimension of the YM coupling constant $[e^2]=M^{-1}
L^{D-4}$ (in $c=1$ units)  is  the same as the dimension of the
physical Newton's constant $[G]=M^{-1} L$.

For the reason just explained, the main body of this paper is
focused on the lowest super-critical dimension $D=5$. In Section~3
we show that in this case there exists a countable family of
regular (by regularity we mean analyticity inside the future light
cone) spherically symmetric self-similar solutions labelled by a
nonnegative integer $n$ (a nodal number). Next, using linear
stability analysis we show in Section~4 that  the number of
unstable modes around a given solution is equal to its nodal
number. The role of self-similar solutions in the dynamical
evolution is studied in Section~5, where we show that: i) the
$n=0$ solution determines a universal asymptotics of singularity
formation for solutions starting from generic "large" initial
data; ii) the $n=1$ solution plays the role of a critical solution
sitting at the threshold of singularity formation. The latter is
in many respects similar to the critical behaviour at the
threshold of black hole formation in gravitational collapse. In
both cases  the threshold of singularity (or black hole) formation
can be identified with the codimension-one stable manifold of a
self-similar solution with exactly one unstable mode. These
similarities are discussed in detail in Section~6.

We consider also the Cauchy problem for the  YM equations in
$D=4$. Despite intensive studies of this borderline case, the
problem of global existence is open. In Section~7 we describe
numerical simulations which, in combination with analytic results,
strongly suggest that large-energy solutions do blow up. We show
that the process of singularity formation is due to concentration
of energy and proceeds via adiabatic shrinking of the instanton
solution. At the end, a recent attempt of determining the
asymptotic rate of shrinking is sketched.

We remark  that there are close parallels between  YM equations in
$D+1$ dimensions and
 wave
maps in $(D-2)+1$ dimensions~\cite{ca_sh_ta}. Indeed, many of the
phenomena described here are mirrored for the equivariant wave
maps into spheres in three~\cite{bi1, bi_ch_ta1} and
two~\cite{bi_ch_ta2} spatial dimensions.

Sections~5 and 7 of this survey are based on joint work with Z.
Tabor~\cite{bi_ta}. The material of Sections~3 and 4 is new.
\section{Setup}
We consider Yang-Mills (YM) fields in $D+1$ dimensional Minkowski
spacetime
 (in the following Latin and Greek
indices take the values $1,2,\dots,D$ and $0,1,2,\dots,D$
respectively).
 The gauge potential $A_{\alpha}$ is
a one-form with values in the Lie algebra $g$ of a compact Lie group $G$.
 In terms of the curvature
$F_{\alpha\beta}=\partial_{\alpha}A_{\beta}-\partial_{\beta}A_{\alpha}+
[A_{\alpha},A_{\beta}]$ the  action is
\begin{equation}\label{action}
 S=\frac{1}{e^2} \int Tr(F_{\alpha\beta} F^{\alpha\beta}) d^D\! x\: dt,
\end{equation}
where $e$ is the gauge coupling constant. Hereafter we  set $e=1$.
 The YM equations derived from (\ref{action})  are
\begin{equation}\label{ymeq}
\partial_{\alpha}F^{\alpha\beta}+  [A_{\alpha}, F^{\alpha\beta}] =
0.
\end{equation}
As written, this equation is underdetermined because of the gauge
invariance
\begin{equation}\label{gauge}
  A_{\alpha} \rightarrow U^{-1} A_{\alpha} U + U^{-1}
  \partial_{\alpha} U,
\end{equation}
where $U$ is an arbitrary function with values in $G$. In order to
correctly formulate the Cauchy problem for equation (\ref{ymeq}),
one must impose additional conditions which fix this gauge
ambiguity. We shall not discuss this issue here because in the
spherically symmetric ansatz, to which this paper is restricted,
the gauge is fixed automatically.

For simplicity, we take here $G=SO(D)$ so the elements of $so(D)$
can be considered as skew-symmetric $D\times D$ matrices and the
Lie bracket is the usual commutator. Assuming
 the spherically symmetric ansatz~\cite{du}
\begin{equation}
A^{ij}_{\mu}(x) = \left(\delta^i_{\mu}x^j-\delta^j_{\mu}x^i\right)
\frac{1-w(t,r)}{r^2},
\end{equation}
the YM equations reduce to the scalar semilinear wave equation for
the magnetic potential $w(t,r)$
\begin{equation}\label{wave}
-w_{tt}+ \Delta_{(D-2)} w + \frac{D-2}{r^2} w (1-w^2) =0,
\end{equation}
where $\Delta_{(D-2)}=\partial^2_{r} +\frac{D-3}{r}\partial_r$ is
the radial Laplacian in $D-2$ dimensions. The central question for
equation (\ref{wave}) is:  can solutions starting from  smooth
initial data
\begin{equation}\label{ics}
  w(0,r) = f(r), \qquad w_t(0,r) = g(r)
\end{equation}
become singular in future?  As mentioned above, in the physical
$D=3$ dimensions Eardley and Moncrief answered this question in
the  negative~\cite{ea_mo}. However, simple heuristic arguments
indicate that the property of global regularity enjoyed by the YM
equations in $D=3$ might break down in higher dimensions. In order
to see why the global behaviour of solutions is expected to depend
critically on the dimension $D$, we recall two basic facts. The
first fact is  the conservation of (positive definite) energy
\begin{equation}\label{energy}
 E = \int_{R^D}
Tr\left(F_{0i}^2 + F_{ij}^2\right) d^D x = c(D)
\int\limits_0^{\infty} \left(w_t^2 +w_r^2 +\frac{D-2}{2 r^2}
(1-w^2)^2\right) r^{D-3} dr,
\end{equation}
where the coefficient $c(D)=(D-1) vol(S^{D-1})$ follows from the
integration over the angles and taking the trace. The second fact
 is scale-invariance of  the YM equations: if
$A_{\alpha}(x)$ is a solution of (\ref{ymeq}), so is $\tilde
A_{\alpha}(x)=\lambda^{-1} A_{\alpha}(x/\lambda)$, or
equivalently, if $w(t,r)$ is a solution of (\ref{wave}), so is
$\tilde w(t,r)=w(t/\lambda,r/\lambda)$. Under this scaling the
energy scales as
 $\tilde E=\lambda^{D-4}
E$, hence the YM equations are subcritical for $D\leq 3$, critical
for $D=4$, and supercritical for $D\geq 5$. In the subcritical
case, shrinking of solutions to arbitrarily small scales costs
infinite amount of  energy, so it is forbidden by energy
conservation. In other words,  transfer of energy to arbitrarily
high frequencies is impossible and consequently  the Cauchy
problem should be well posed in the energy norm. This important
fact was proved in $D=3$ by Klainerman and Machedon~\cite{kl_ma},
who thereby strengthened the result of Eardley and Moncrief. In
the supercritical case, shrinking of solutions might be
energetically favourable and consequently singularities are
anticipated. In fact, we shall show below  that singularities do
form already in the lowest supercritical dimension $D=5$. In the
critical dimension $D=4$ the problem of singularity formation
 is more subtle because the scaling argument is inconclusive.
\section{Self-similar solutions in $D=5$}
In order to set the stage for the discussion of singularity
formation we first need to analyze in detail the structure of
self-similar solutions of equation (\ref{wave}). As we shall see,
these solutions play a key role in understanding the nature of
blowup. By definition, self-similar solutions are invariant under
dilations $w(t,r) \rightarrow w(t/\lambda,r/\lambda)$, hence they
have the form
\begin{equation}\label{ssansatz}
w(t,r)=W(\eta), \quad \quad \eta=\frac{r}{T-t},
\end{equation}
where  a positive constant $T$, clearly allowed by the time
translation invariance, is introduced for later convenience.  Note
that for a self-similar solution we have
\begin{equation}\label{ssblow}
  \partial_r^2 W(\eta)\Bigr\rvert_{r=0} = \frac{1}{(T-t)^2} W''(0),
\end{equation}
hence the solution becomes singular at the center when $t
\rightarrow T$ (there is no blowup in the first derivative because
regularity demands that $W'(0)=0$). Thus, each self-similar
solution $W(\eta)$ provides  an explicit example of a singularity
developing in finite time from smooth initial data.

Substituting the ansatz (\ref{ssansatz}) into (\ref{wave}) one
obtains the ordinary differential equation
\begin{equation}\label{sseq}
 W''+\left(\frac{D-3}{\eta}
+\frac{(D-5)\eta}{1-\eta^2}\right) W' +\frac{D-2}{\eta^2
(1-\eta^2)} W(1-W^2) = 0. \end{equation}
 As explained in the introduction, because of the expected connections with the Einstein
 equations,
we are mainly interested in the lowest super-critical dimension
$D=5$. In this case equation  (\ref{sseq}) reduces to
\begin{equation}\label{sseq5}
 W''+ \frac{2}{\eta} W'
+\frac{3}{\eta^2 (1-\eta^2)} W (1-W^2) = 0.
\end{equation}
Although the similarity coordinate $\eta$ is  natural  in the
discussion of singularity formation, it has a disadvantage of not
covering the region $t>T$, in particular it does not extend to the
future light cone of the point $(T,0)$. For this reason we define
  a new coordinate $x=1/\eta$ which
 covers the whole spacetime:  the past and the future light cones are located at
 $x=1$ and $x=-1$, respectively; while the center $r=0$ corresponds to
 $x=\infty$ (for $t<T$) and $x=-\infty$ (for $t>T$).
In terms of $x$ equation (\ref{sseq5}) becomes
 \begin{equation}\label{main}
  (x^2-1) W''  +  3 W (1-W^2) =0.
\end{equation}
We first consider this equation inside the past light cone, that
is for $1 \leq x < \infty$ and impose the  boundary conditions
\begin{equation}\label{bc}
 W(1)=0
\quad
 \mbox{and} \quad W(\infty)=\pm 1,
\end{equation}
which follow from  the demand of smoothness at the endpoints. As
we shall see below, once a solution to this boundary value problem
is constructed, its extension  beyond the past light cone can be
easily done.

 To show that  equation (\ref{main}) admits solutions satisfying
  (\ref{bc})
 we shall
employ a shooting technique. The main idea of this method is to
replace the  boundary value problem by the  initial value problem
with  initial data imposed at one of the endpoints and then
adjusting these data so that the solution hits the desired
boundary value at the second endpoint. In the case at hand we
shall shoot from $x=1$ towards infinity.
 Substituting  a formal power series expansion about $x=1$ into
 (\ref{main}) one finds
 the asymptotic behaviour
 \begin{equation}\label{aorbit}
 W(x) = a (x-1)
-\frac{3 a}{4} (x-1)^2 + O\left((x-1)^3\right),
\end{equation}
where $a$ is a free parameter determining uniquely the whole
series. In the following a solution of equation (\ref{main})
starting at $x=1$ with the asymptotic behaviour (\ref{aorbit})
will be called an $a$-orbit. Without loss of generality we may
assume that $a \geq 0$. We claim  that there is a countable set of
values $\{a_n\}$ for which the $a_n$-orbits exist for all $x \geq
1$ and have the desired asymptotics at infinity (such orbits will
be called connecting). The proof consists of several steps.
 \vskip 0.2cm \noindent
\emph{Step~1 (Local existence)}. First, we  need to show that
$a$-orbits do in fact exist, that is, the series (\ref{aorbit})
has a nonzero radius of convergence. Since the point $x=1$ is
singular, this fact does not follow from standard theorems.
Fortunately, in~\cite{bfm} Breitelohner, Forg\'acs, and Maison
have derived the following result concerning the behaviour of
solutions of a system of ordinary differential equations near a
singular point: \vskip 0.1cm \noindent
\emph{Theorem [BFM].
 Consider a system of first order differential
equations for $n+m$ functions $u=(u_1,...,u_n)$ and
$v=(v_1,...,v_m)$
\begin{equation}\label{bfm}
y\frac{du_i}{dy} = y^{\mu_i} f_i(y,u,v),\qquad y\frac{dv_i}{dy} =
-\lambda_i v_i + y^{\nu_i} g_i(y,u,v),
\end{equation}
where constants $\lambda_i>0$ and integers $\mu_i, \nu_i\geq1$ and
let $C$ be an open subset of $R^n$ such that the functions $f$ and
$g$ are analytic in the neighbourhood of $y=0, u=c, v=0$ for all
$c\in C$. Then there exists an $n$-parameter family of solutions
of the system (\ref{bfm}) such that
\begin{equation}\label{bfm2}
  u_i(y) = c_i + O(y^{\mu_i}), \qquad v_i(y)=O(y^{\nu_i}),
\end{equation}
where $u_i(y)$ and $v_i(y)$ are defined for all $c\in C,
|y|<y_0(c)$ and are analytic in $y$ and $c$.} \vskip 0.1cm

 We
shall make use of this theorem to prove the local existence of
$a$-orbits. In order
  to put equation (\ref{main}) into the form (\ref{bfm2}) we define
  the variables
\begin{equation}\label{ch1}
 y=x-1, \quad  u(y)= W', \quad v(y)=\frac{W}{x-1}-W',
\end{equation}
and get
\begin{equation}
 y v' = -v + y f, \qquad y u' = y f, \qquad
f=\frac{3 (u+v) \left[1-y^2 (u+v)^2\right]}{2+y}.
\end{equation}
Since the function $f(y,u,v)$ is analytic near $y=0$ for any $u$
and $v$, according to the BFM theorem there exists a one-parameter
family of local solutions such
 that
\begin{equation}\label{loc}
 u(y)= a  + O(y), \qquad v(y)= O(y),
\end{equation}
Transforming (\ref{loc}) back to the original variables we obtain
the
 behaviour of $a$-orbits.
 \vskip 0.2cm \noindent
\emph{Step~2 (A priori global behaviour)}.  It follows immediately
from (\ref{main}) that  for $x>1$ a solution cannot have a maximum
(resp. minimum) for $W>1$ (resp. $W<-1$). Thus, once the solution
leaves the strip $|W|<1$, it cannot reenter it (actually, such a
solution becomes singular for a finite $x$). It is also clear that
as long as $|W|<1$ the solution cannot go singular. To derive the
asymptotics at infinity of $a$-orbits that stay in the strip
$|W|<1$  we shall make use of the following functional
 \begin{equation}\label{Q}
Q(x) = \frac{1}{2} (x^2-1) {W'}^2 -\frac{3}{4} (1-W^2)^2.
\end{equation}
 For solutions of equation (\ref{main}) we have
\begin{equation}\label{Q'}
 Q'(x)= x {W'}^2,
 \end{equation}
so $Q(x)$ is monotone increasing.
 Now, we shall show that solutions satisfying $|W|<1$ for
all $x \geq 1$ tend  to $W=\pm 1$ as $x\rightarrow \infty$. To see
this, first notice that for such solutions $Q$ must be negative
because if $Q(x_0)>0$ for some $x_0>1$ then $|W'|$ is strictly
positive for $x>x_0$ so the solution  must leave the strip $|W|<1$
in finite time. Since $Q' \geq 0$ and $Q \leq 0$, it follows that
$Q$ has a nonpositive limit at infinity which in turn implies by
(\ref{Q'}) that $\lim_{x \rightarrow \infty} x W' =0$ and by
(\ref{Q}) that $\lim_{x \rightarrow \infty} W$ exists. By
L'H\^{o}pital's rule we have  $\lim_{x \rightarrow \infty} x^2 W''
=0$ and using (\ref{main}) again, we get that $\lim_{x \rightarrow
\infty} W$ equals $\pm 1$ or $0$. The latter is impossible because
then $Q(\infty)=-3/4$ is a global minimum  contradicting the fact
that $Q$ increases. Thus, $W(\infty) =\pm 1$.
 \vskip 0.2cm
\noindent\emph{Step~3} (i) (\emph{Behaviour of $a$-orbits for
small $a$}).
 Rescaling $w(x)=W(x)/a$ we get
 \begin{equation}\label{resc}
 (x^2-1) w'' + 3 w (1-a^2 w^2) =0, \qquad  w(1)=0,\quad w'(1)=1.
 \end{equation}
 As $a \rightarrow 0$, the solutions of this
  equation  tend
uniformly on compact intervals to the solution of the limiting
equation
\begin{equation}\label{bla1}
 (x^2-1) w'' + 3 w =0
 \end{equation}
  with the same initial condition. This equation can be solved
  explicitly but for the purpose of the argument
   it suffices  to notice that its solution, call it $w_L(x)$, is
  oscillating at infinity. Since $W(x,a) \approx a w_L(x)$   up to an arbitrarily
  large $x$ if $a$ is sufficiently small, it follows that
 the number of zeros of the solution $W(x,a)$ tends to infinity as
 $a \rightarrow 0$.
\vskip 0.1cm \noindent (ii)  (\emph{Behaviour of $a$-orbits for
large $a$}.) We rescale the variables, setting $y=a (x-1)$,
   $\bar w(y)=W(x)$ to get
  \begin{equation}\label{resc2}
  \bar y (y+ 2 a) \bar w'' + 3 \bar w (1-\bar w^2)  =0, \qquad \bar w(0)=0, \quad \bar
  w'(0)=1.
  \end{equation}
As $a \rightarrow \infty$, the solutions of this equation tend
uniformly on compact intervals to the solution $\bar w(y)=y$ of
the limiting equation $\bar w''=0$. Thus,  $W(x,a) \approx  a
(x-1)$ for large $a$ and therefore the $a$-orbit  crosses $W=1$
for a finite $x$.
\vskip 0.2cm \noindent\emph{Step~4 (Shooting argument)}.
 We define the set
\begin{equation}\label{bla2}
 A_0 =\{a\,|\, W(x,a)\,\,  \mbox{strictly increases up
 to some}
  \,\, x_0 \,\,\mbox{where}\,\, W(x_0,a)=1\}.
\end{equation}
We know from Step~3 that the set $A_0$ is nonempty (because the
$a$-orbits with large $a$ belong to it) and bounded below (because
the $a$-orbits with small $a$ do not belong to it).
 Thus $a_0=\inf A_0$ exists.
The solution $W(x,a_0)$  cannot cross the line $W=1$ at a finite
$x$ because the same would be true for nearby solutions, violating
the definition of $a_0$. Thus, $0\leq W(x,a_0)<1$ for all $x$ and
hence, by Step~2, $\lim_{x\rightarrow \infty} W(x,a_0) =1$. This
completes the proof of existence of the nodeless self-similar
solution $W_0(x)\stackrel{def}{=} W(x,a_0)$.

Next, let us consider the solution with $a=a_0-\epsilon$ for small
$\epsilon>0$.   By the definition of $a_0$ there must be a point
 $x_0$ where this solution
attains a positive local maximum $W(x_0)<1$ and since no minima
are possible for $0<W<1$, it follows that there must be a point
$x_1>x_0$ where $W(x_1,b)=0$.
 We shall show that $Q(x_1,a)>0$ provided that $\epsilon$ is
sufficiently small. As argued above this would imply  that the
solution $W(x,a)$ leaves the strip $|W|<1$ via $W=-1$.
 From (\ref{Q'}) we have
  \begin{equation}\label{bla1}
   Q(x_1)-Q(x_0)=\int\limits_{x_0}^{x_1} x {W'}^2 dx = -\int\limits_0^{W(x_0)} x W' dW.
\end{equation}
In order to estimate
the last integral note that for $x>x_0$
\begin{equation}
Q(x)-Q(x_0)=\frac{1}{2} (x^2-1) {W'}^2 - \frac{3}{4} (1-W^2)^2 +
\frac{3}{4} (1-W^2(x_0))^2
> 0,
\end{equation}
 so $x |W'|>\sqrt{\frac{3}{2}} \sqrt{(1-W^2)^2 - (1-W^2(x_0))^2}$ . Substituting
  this into (\ref{bla1}) one gets
\begin{equation}
 Q(x_1)>-\frac{3}{4} \left(1-W^2(x_0)\right)^2 + \sqrt{\frac{3}{2}}
\int\limits_0^{W(x_0)} \sqrt{(1-W^2)^2 - (1-W^2(x_0))^2 }\, dW.
\end{equation}
The right hand side of this inequality is equal to $\sqrt{2/3}$
for $W(x_0)=1$ so, by continuity, it remains strictly positive for
$W(x_0)$ near $1$. By taking a sufficiently small $\epsilon$ we
can have $W(x_0)$ arbitrarily close to $1$, hence $Q(x_1)>0$ which
proves that $a$-orbits with $a=a_0-\epsilon$ have exactly one
zero. This means that the set $A_1=\{a\,|\,
  W(x,a)$ increases up to some $x_0$ where it attains a positive local
  maximum $W(x_0)<1$ and then decreases monotonically up to some
  $x_1$ where $W(x_1)=-1\}$ is nonempty.  Let $a_1=\inf A_1$.
  By Step~3, $a_1$ exists and  is strictly positive. Using the same
 argument
as above we conclude that the $a_1$-orbit must stay in the region
 $|W|<1$ for all $x$, hence $\lim_{x\rightarrow \infty}
 W(x,a_1)=-1$. This completes the proof of existence of the self-similar solution
 $W_1(x)\stackrel{def}{=}W(x,a_1)$ with exactly one zero.

The subsequent connecting orbits are obtained by induction.  We
conclude that there exists a countable family of self-similar
solutions $W_n(x)$ indexed  by the integer $n=0,1,...$  $n$ which
counts the number of zeros for $x>1$. \vskip 0.15cm \noindent
\emph{Remark.} Since the sequence $\{a_n\}$ is decreasing and
bounded below by zero, it has a nonnegative limit
$\lim_{n\rightarrow\infty}a_n=a^*\geq 0$. If $a^*>0$, then the $
a^*$-orbit cannot leave the region $|W|<1$ for a finite $x$
(because the set of such orbits is clearly open) hence it must be
a connecting orbit with some finite number of zeros. But this
contradicts the fact that the number of zeros of $a_n$-orbits
increases with $n$. Hence, $a^*=0$. This implies that for any
finite $x$, $W_n(x)$ goes to zero when $n\rightarrow \infty$.
\vskip 0.15cm
 We remark that the existence of the solution $W_0$ was
 first shown by Cazenave, Shatah, and Tahvildar-Zadeh~\cite{ca_sh_ta}
 via a variational method.
 \vskip 0.2cm
 \noindent The shooting technique  is not only a powerful
 analytical tool; it is also an efficient numerical method of solving
 two-point boundary value problems.  The numerical results produced by this method
 are
 shown in Table~1 and Figure~1.
\begin{figure}[h]
\centering
\includegraphics[width=0.8\textwidth]{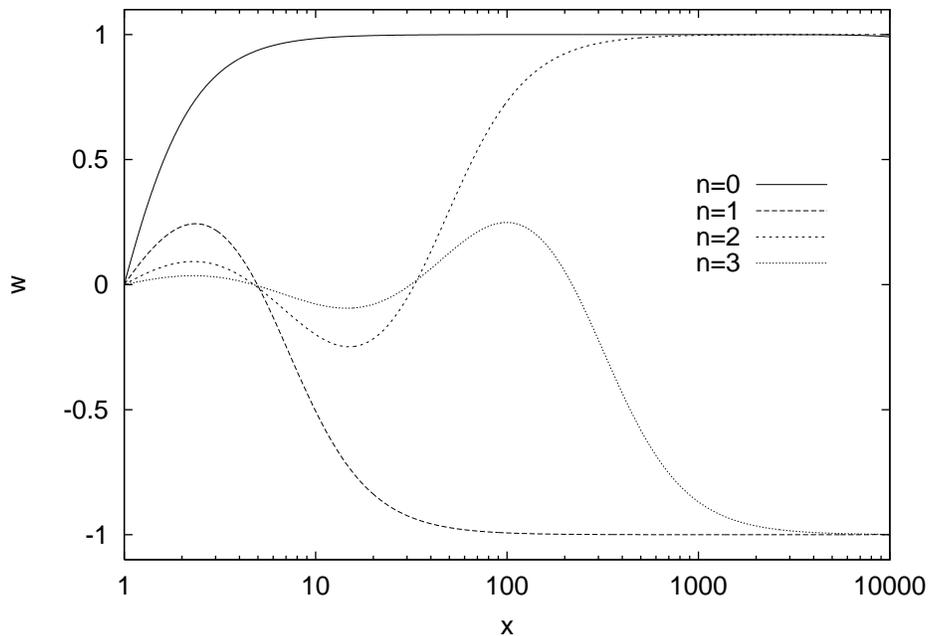}
\caption{\small{The first four self-similar solutions $W_n(x)$.}
}\label{fig1}
\end{figure}
  \begin{table}[h]
\centering
$$
\begin{tabular}{|c|c|c|c|c|c|c|} \hline
$n$ & 0 & 1 & 2 & 3 & 4 & 5\\ \hline $a_n$ & 1.25 & 0.4813158 &
0.1864517 &
0.0722966 & 0.02803703 & 0.01087315\\
\hline
\end{tabular}
$$
\caption{The shooting parameters of solutions $W_n$ for $n\leq
5$.}
\end{table}

Surprisingly,  it turned out that $a_0=5/4$
 (with very good accuracy). This was a hint that the solution $W_0$
 has a simple closed form. Indeed,  playing  with the power series
 expansion (\ref{aorbit}) we found that
\begin{equation}\label{w0}
W_0(x)=\frac{x^2-1}{x^2+\frac{3}{5}}.
 \end{equation}
Below we show an amusing calculation by Maple which helped us in
  finding this formula. \vskip 0.2cm
\def\emptyline{\vspace{12pt}}
\DefineParaStyle{Maple Output} \DefineCharStyle{2D Math}
\DefineCharStyle{2D Output}
\begin{maplegroup}
\begin{mapleinput}
\mapleinline{active}{1d}{restart;}{%
}
\end{mapleinput}

\end{maplegroup}
\begin{maplegroup}
\begin{mapleinput}
\mapleinline{active}{1d}{with(DEtools):}{%
}
\end{mapleinput}

\end{maplegroup}
\begin{maplegroup}
\begin{mapleinput}
\mapleinline{active}{1d}{with(numapprox):}{%
}
\end{mapleinput}

\end{maplegroup}
\begin{maplegroup}
\begin{mapleinput}
\mapleinline{active}{1d}{ode:=(x^2-1)*diff(w(x),x$2)+3*w(x)*(1-w(x)^2)=0;}{%
}
\end{mapleinput}

\mapleresult
\begin{maplelatex}
\mapleinline{inert}{2d}{ode := (x^2-1)*diff(w(x),`$`(x,2))+3*w(x)*(1-w(x)^2) = 0;}{%
\[
\mathit{ode} := (x^{2} - 1)\,({\frac {\partial ^{2}}{\partial x^{
2}}}\,\mathrm{w}(x)) + 3\,\mathrm{w}(x)\,(1 - \mathrm{w}(x)^{2})=
0
\]
}
\end{maplelatex}

\end{maplegroup}
\begin{maplegroup}
\begin{mapleinput}
\mapleinline{active}{1d}{ic:=w(1)=0,D(w)(1)=5/4;}{%
}
\end{mapleinput}

\mapleresult
\begin{maplelatex}
\mapleinline{inert}{2d}{ic := w(1) = 0, D(w)(1) = 5/4;}{%
\[
\mathit{ic} := \mathrm{w}(1)=0, \,\mathrm{D}(w)(1)= {\displaystyle
\frac {5}{4}}
\]
}
\end{maplelatex}

\end{maplegroup}
\begin{maplegroup}
\begin{mapleinput}
\mapleinline{active}{1d}{sol:=dsolve(\{ode,ic\},w(x));}{%
}
\end{mapleinput}

\mapleresult
\begin{maplelatex}
\mapleinline{inert}{2d}{sol := ;}{%
\[
\mathit{sol} :=
\]
}
\end{maplelatex}

\end{maplegroup}
\begin{maplegroup}
\begin{mapleinput}
\mapleinline{active}{1d}{sol_formal:=rhs(dsolve(\{ode,ic\},w(x),type=series));}{%
}
\end{mapleinput}

\mapleresult
\begin{maplelatex}
\mapleinline{inert}{2d}{sol_formal :=
series(5/4*(x-1)-15/16*(x-1)^2+25/64*(x-1)^3+25/256*(x-1)^4-375/1024*(
x-1)^5+O((x-1)^6),x=-(-1),6);}{%
\maplemultiline{
\mathit{sol\_formal} :=  \\
{\displaystyle \frac {5}{4}} \,(x - 1) - {\displaystyle \frac {15
}{16}} \,(x - 1)^{2} + {\displaystyle \frac {25}{64}} \,(x - 1)^{
3} + {\displaystyle \frac {25}{256}} \,(x - 1)^{4} -
{\displaystyle \frac {375}{1024}} \,(x - 1)^{5} + \mathrm{O}((x
 - 1)^{6}) }
}
\end{maplelatex}

\end{maplegroup}
\begin{maplegroup}
\begin{mapleinput}
\mapleinline{active}{1d}{pade_sol:=pade(sol_formal,x=1,[2,2]);}{%
}
\end{mapleinput}

\mapleresult
\begin{maplelatex}
\mapleinline{inert}{2d}{pade_sol := (5/8*(x-1)^2+5/4*x-5/4)/(-1/4+5/4*x+5/8*(x-1)^2);}{%
\[
\mathit{pade\_sol} := {\displaystyle \frac {{\displaystyle \frac
{5}{8}} \,(x - 1)^{2} + {\displaystyle \frac {5}{4}} \,x -
{\displaystyle \frac {5}{4}} }{ - {\displaystyle \frac {1}{4}}
 + {\displaystyle \frac {5}{4}} \,x + {\displaystyle \frac {5}{8}
} \,(x - 1)^{2}}}
\]
}
\end{maplelatex}

\end{maplegroup}
\begin{maplettyout}
\end{maplettyout}

\begin{maplegroup}
\begin{mapleinput}
\mapleinline{active}{1d}{sol:=simplify(pade_sol);}{%
}
\end{mapleinput}

\mapleresult
\begin{maplelatex}
\mapleinline{inert}{2d}{sol := 5*(x^2-1)/(3+5*x^2);}{%
\[
\mathit{sol} := 5\,{\displaystyle \frac {x^{2} - 1}{3 + 5\,x^{2}}
}
\]
}
\end{maplelatex}

\end{maplegroup}
\begin{maplegroup}
\begin{mapleinput}
\mapleinline{active}{1d}{subs(w(x)=sol,ode);}{%
}
\end{mapleinput}

\mapleresult
\begin{maplelatex}
\mapleinline{inert}{2d}{(x^2-1)*diff(5*(x^2-1)/(3+5*x^2),`$`(x,2))+15*(x^2-1)*(1-25*(x^2-1)^2
/((3+5*x^2)^2))/(3+5*x^2) = 0;}{%
\[
(x^{2} - 1)\,({\frac {\partial ^{2}}{\partial x^{2}}}\,(5\,
{\displaystyle \frac {x^{2} - 1}{3 + 5\,x^{2}}} )) + 15\,
{\displaystyle \frac {(x^{2} - 1)\,(1 - 25\,{\displaystyle \frac
{(x^{2} - 1)^{2}}{(3 + 5\,x^{2})^{2}}} )}{3 + 5\,x^{2}}} =0
\]
}
\end{maplelatex}

\end{maplegroup}
\begin{maplegroup}
\begin{mapleinput}
\mapleinline{active}{1d}{simplify(\%);}{%
}
\end{mapleinput}

\mapleresult
\begin{maplelatex}
\mapleinline{inert}{2d}{0 = 0;}{%
\[
0=0
\]
}
\end{maplelatex}

\end{maplegroup}
\begin{maplegroup}
\begin{mapleinput}
\end{mapleinput}

\end{maplegroup}
So far our analysis of self-similar solutions was restricted to
the interior of the past light cone of the singularity. To show
that the solutions $W_n$ represent genuine naked singularities, we
need to extend them to the future light cone, that is to $x=-1$.
Fortunately, such an extension creates no problem because an
$a$-orbit shot backwards from $x=1$ cannot go singular before
reaching $x=-1$. This follows immediately  from (\ref{main}) by
observing that, in the interval  $-1<x<1$, $W(x)$ is concave down
(resp. up) for $W>1$ (resp. $W<-1$), hence $W(x)$ remains bounded
as $x \rightarrow -1^+$. Moreover, the function $Q(x)$ is negative
and decreasing near $x=-1$, thus $\lim_{x \rightarrow -1^+} Q(x)$
exists which implies in turn that $c=\lim_{x \rightarrow
-1^+}W(x)$  exists. Having that, the standard asymptotic analysis
gives the following leading order behaviour for $x \rightarrow
-1^+$
\begin{equation}\label{asym4}
W(x) \sim c + \frac{3}{2} c (1-c^2) (x+1) \ln(x+1).
\end{equation}
The singular logarithmic term in (\ref{asym4}) can be eliminated
by fine-tuning the shooting parameter $a$,  however this is not
expected to happen for the solutions $W_n(x)$ because in their
construction the freedom of adjusting  $a$ was already used to
tune away the singular behaviour for $x>1$. We conclude that the
self-similar solutions $W_n$ are $C^0$ at the future light cone
and are analytic everywhere below it. The only (somewhat
surprising) exception is the solution $W_0$ which is analytic in
the entire spacetime.
\section{Linear stability of self-similar solutions}
In this section we study the linear stability of self-similar
solutions $W_n$. This analysis is essential in determining  the
role of self-similar solutions in dynamics.
 We restrict
attention to the interior of the past light cone of the
 point $(T,0)$ and define the new time coordinate
 $s=-\ln{\sqrt{(T-t)^2-r^2}}$. Note that $s\rightarrow \infty$ when $t \rightarrow T$, and
 the lines of constant $s$ are orthogonal to the rays of constant $x$.
In terms of $s$ and $x$, equation (\ref{wave}) becomes (for $D=5$)
\begin{equation}\label{mains}
-\frac{e^{2 s}}{x^2-1}  (e^{-2 s} w_{s})_{s} + (x^2-1) w_{xx} + 3
w(1-w^2) =0.
 \end{equation}
  Of course, this equation reduces to
(\ref{main}) if $w$ does not depend on $s$. In order to determine
the stability of self-similar solutions $W_n$ we seek solutions of
(\ref{mains}) in the form $w(s,x)=W_n(x)+ v(s,x)$. Neglecting the
$O(v^2)$ terms we obtain the linear evolution equation for the
perturbation $v(s,x)$
\begin{equation}\label{linper}
-\frac{e^{2 s}}{x^2-1}  (e^{-2 s} v_{s})_{s} + (x^2-1) v_{xx} +
3(1-3 W_n^2) v =0.
\end{equation}
Substituting $v(s,x)=e^{(\alpha+1) s} \sqrt{x^2-1} \; u(x)$ into
(\ref{linper})  we get the  eigenvalue problem in the standard
Sturm-Liouville form
\begin{equation}\label{s1}
-\frac{d}{dx}\left((x^2-1) \frac{du}{dx}\right) - 3 (1- 3 W^2_n) u
= \frac{\lambda}{x^2-1} u,
\end{equation}
 where $\lambda=-\alpha^2$. Using
the variable $\rho=\frac{1}{2}\ln(\frac{x-1}{x+1})$ ranging  from
zero to infinity we transform (\ref{s1}) into  the radial
Schr\"odinger equation
\begin{equation}\label{schr}
 -\frac{d^2 u}{d \rho^2}  + V_n u = \lambda u, \qquad
V_n=-\frac{3 (1-3 W_n^2)}{\sinh^2{\!\rho}}.
 \end{equation}
The potential $V_n(\rho)$ has a typical "quantum mechanical" shape
(see Figure~2) with the asymptotics
 \begin{equation}\label{pot}
 V_n(\rho) \sim \left\{
  \begin{array}{rl}
    6/\rho^2  & \,\,\mbox{for} \,\,\,\rho \rightarrow 0, \\
    -12 \exp(-2 \rho) & \,\,\mbox{for} \,\,\,\rho \rightarrow \infty.
  \end{array}
\right.
\end{equation}
Note that the potential can be expressed in the form $V_n(\rho)=l
(l+1)/\rho^2+V_n^{reg}(\rho)$ with $l=2$, where the regular part
$V_n^{reg}(\rho)$ is everywhere  negative and $V_n^{reg}(0)
\rightarrow -\infty$ as $n\rightarrow \infty$.
\begin{figure}[h]
\centering
\includegraphics[width=0.8\textwidth]{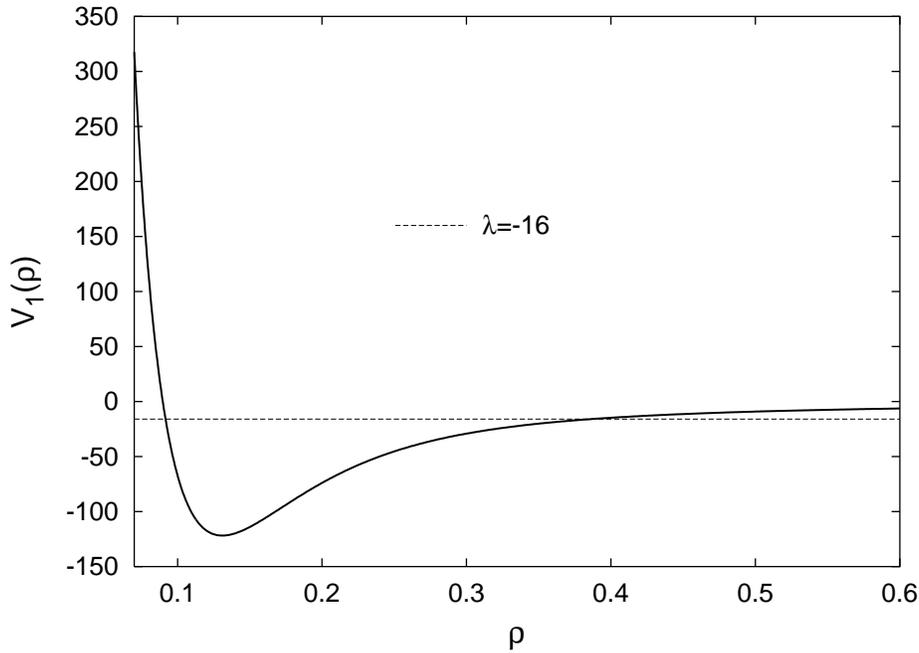}
\caption{\small{The potential for the perturbations around the
self-similar solution $W_1$. The single bound state with energy
$\lambda=-16$ is indicated.}}\label{fig2}
\end{figure}

\noindent Both endpoints $\rho=0$ and $\rho=\infty$ are of the
limit-point type, that is, exactly one solution near each point is
square-integrable (admissible). Near $\rho=0$ the admissible
solutions behave as $u(\rho) \sim \rho^3$. For $\rho \rightarrow
\infty$ and $\lambda<0$ the admissible solutions behave as
 $u(\rho) \sim e^{-\alpha \rho}$ (recall that $\alpha=\sqrt{-\lambda}$). All
  $\lambda \geq 0$ belong to the continuous spectrum.

Let $u^n_k$ (resp. $\alpha^n_k$) denote the $k$th eigenfunction
(resp. eigenvalue) about the solution $W_n$. The numerically
generated spectra are shown in Table~2.
\begin{table}[h]
\centering
$$
\begin{tabular}{|c|c|c|c|c|c|} \hline
$n$ & $\alpha^n_0$ & $\alpha^n_1$ & $\alpha^n_2$ & $\alpha^n_3$ & $\alpha^n_4 $\\
 \hline
0 & 0 &  &  &  &\\
1 & 0 & 4 &  & &\\
2 & 0 & 4  & 27.407 & &\\
3 & 0  & 4 & 27.379 & 182.49 &\\
4 & 0 & 4 & 27.374 & 182.18 & 1214.5\\
... & ...&... &... &... & ...\\
$\infty$ & 0 & 4 & 27.37319 & 182.1202 & 1210.917\\
 \hline
\end{tabular}
$$
\caption{The eigenvalues of the perturbations about the  first
five solutions $W_n$ obtained numerically. The pseudo-eigenvalue
$\alpha=0$ is also included. The last row corresponding to
$n=\infty$ was obtained by solving numerically  the transcendental
equation (\ref{quant}).}
\end{table}

\noindent We point out that although $\alpha=0$ is not a genuine
eigenvalue, it is distinguished from the strictly positive part of
the continuous spectrum by the fact that  the corresponding
non-square-integrable pseudo-eigenfunction, called the zero mode,
is subdominant at infinity. The existence of the zero mode is due
to the time translation symmetry, or in other words, the freedom
of shifting the blowup time $T$ in (\ref{ssansatz}). To see this,
consider the self-similar solution with a shifted blowup time
$W_n((T'-t)/r)$, where $T'=T+\epsilon$. In terms of the original
similarity variables $s=-\ln{\sqrt{(T-t)^2-r^2}}$ and $x=(T-t)/r$,
we have
\begin{equation}
W_n\!\left(\frac{T'-t}{r}\right)=W_n(x+\epsilon e^{s}
\sqrt{x^2-1}) = W_n(x) + \epsilon e^{s} \sqrt{x^2-1} \;W'_n(x) +
O(\epsilon^2),
\end{equation}
hence the  perturbation generated by shifting the blowup time
corresponds to $\alpha=0$ and  has the form
\begin{equation}\label{zeromode}
u^n_0=  \sqrt{x^2-1} \;W'_n(x) = \sinh^2{\!\rho}\; W'_n(\rho).
\end{equation}
An alternative way of deriving   this result is to take
$(\sinh^2{\!\rho}\: W')' + 3 W (1-W^2)=0$, which is (\ref{main})
reexpressed in terms of $\rho$, differentiate it and compare with
(\ref{schr}).

 Since by construction the solution $W_n(\rho)$ has $n$
extrema, it follows from (\ref{zeromode}) that the zero mode
$u^n_0(\rho)$ has $n$ nodes. This implies, by the standard result
from Sturm-Liouville theory, that the potential $V_n$  has exactly
$n$ negative eigenvalues, in agreement with the numerical results
shown in Table~2. We conclude that the self-similar solution $W_n$
has exactly $n$ unstable modes (apart from the unphysical zero
mode). In particular, the fundamental solution $W_0$ is linearly
stable, which makes it a candidate for the attractor.

The rest of this section is a digression concerning a striking
regularity which an acute reader might have already noticed  in
Table~2. Namely, the third column  of Table~2 indicates that for
each $n>0$ the first eigenvalue below the continuous spectrum
$\lambda_1^n=-(\alpha_1^n)^2$ is equal to $-16$ (with the
numerical accuracy of ten decimal places)! This puzzling numerical
fact is calling for an explanation. Clearly, it has something to
do with the particular form of the nonlinearity since, for
instance,
 the analogous problem for  self-similar wave maps from $3+1$
dimensional Minkowski spacetime into the 3-sphere does not have
this property~\cite{bi1}. We suspect that the problem has some
hidden symmetry, yet we cannot exclude  a possibility that the
numerics is misleading
 and  the eigenvalues $\lambda_1^n$ are not precisely equal but their splitting
 is
  beyond the numerical resolution. Some insight into this puzzle can be gained
by analyzing the
 limiting case $n \rightarrow \infty$.
Recall that  $W_n(\rho)$ tends to zero for any $\rho>0$ as $n
\rightarrow \infty$, hence the sequence of potentials $V_n$ has
the following nonuniform limit
\begin{equation}\label{vinf}
 \lim_{n \rightarrow \infty} V_n(\rho) = V_{\infty}(\rho) = -\frac{3}{\sinh^2{\!\rho}}.
 \end{equation}
For the limiting potential $V_{\infty}$ the Schr\"odinger equation
(\ref{schr}) can be solved exactly.  The   solution that is
admissible  at infinity (which as before is the limit-point) is
given by the associated Legendre function of the first kind
\begin{equation}\label{uinf}
u(\rho) = P_{\nu}^{\alpha}(\coth{\rho}), \qquad
\nu=-1/2+i\frac{\sqrt{11}}{2}.
\end{equation}
Here $\nu$ is one of the roots of $\nu(\nu+1)=-3$; the second root
gives the same solution because
 $P_{-1/2+i\beta}^{\alpha}(x)$
with real $\beta$ is real and $P_{-1/2+i\beta}^{\alpha}(x)
=P_{-1/2-i\beta}^{\alpha}(x)$.

As $\rho \rightarrow 0$, the solution (\ref{uinf}) behaves as
\begin{equation}\label{uzero}
  u(\rho) \propto \rho^{\frac{1}{2}} \sin\left(\frac{\sqrt{11}}{2}
  \ln{\rho} + \delta(\alpha)\right),
\end{equation}
so it is always admissible, independently of $\alpha$. This means
that $\rho=0$ is the limit-circle
 point and therefore
in order to have a well-defined self-adjoint problem we need to
impose an additional  boundary condition. In the language of
spectral theory such a condition is called a self-adjoint
extension. The continuous part of the spectrum is the same for all
self-adjoint extensions  but the eigenvalues do
 depend on the choice. In our case the self-adjoint extension
 amounts to fixing $\delta(\alpha)$ - the phase of oscillations of the eigenfunctions
for $\rho \rightarrow 0$.
 The natural choice is to require that the eigenfunctions  oscillate with
the same phase as the zero mode, that is
$\delta(\alpha)=\delta(0)$, or equivalently
\begin{equation}\label{exten}
\lim_{\rho \rightarrow 0} \left\{P_{\nu}^0(\coth{\rho}) u'(\rho) -
{P_{\nu}^0}'(\coth{\rho}) u(\rho)\right\}=0.
\end{equation}
Note that under this condition  the following diagram commutes
\[
\begin{diagram}
\node{V_n} \arrow{e,t}{} \arrow{s}
\node{\{\alpha^n_k\}} \arrow{s,..} \\
\node{V_{\infty}} \arrow{e,t}{} \node{\{\alpha^{\infty}_k\}}
\end{diagram}
\]
 Substituting (\ref{uinf}) into (\ref{exten}) and using the asymptotic expansion
 (for real $\beta$)
 \begin{equation}
 P_{-1/2+i\beta}^{\alpha} (\coth{\rho})\sim \frac{2^{i\beta} \Gamma(i\beta)}
 {\sqrt{2\pi} \:\Gamma(1/2+i\beta-\alpha)} \:\rho^{\frac{1}{2}+i\beta}
 + c.c. \quad \mbox{for} \quad \rho \rightarrow 0,
 \end{equation}
 we obtain the quantization condition for the eigenvalues
 \begin{equation}\label{quant}
 \arg\left\{\Gamma(\frac{1}{2}-i\frac{\sqrt{11}}{2})\:
 \Gamma(\frac{1}{2}+i\frac{\sqrt{11}}{2}+\alpha_k)\right\} = k \pi, \quad k \in \bf{N}.
 \end{equation}
 This transcendental equation has infinitely many roots which for $k\geq 2$ can be obtained only numerically
 (see the last row in Table~2). However, for $k=1$ the exact solution is $\alpha_1=4$
 because
 (accidentally?)
 $\Gamma(1/2+i\sqrt{11}/2+4)=-45 \:\Gamma(1/2+i\sqrt{11}/2)$, as can be readily verified using four times the
 identity $\Gamma(z+1)=z\Gamma(z)$. Thus, we showed that the least negative eigenvalue of the
 limiting potential is equal to
 $-16$.
Although this analysis does
 not resolve the original  puzzle why all $\alpha_1^n$ are equal
 to $4$, it shows at least that $4$ is the accumulation point of
 this sequence.

 The asymptotic distribution of eigenvalues for $k \rightarrow \infty$ can be derived from
 (\ref{quant}) by
 using
 the formula for the asymptotic behaviour of the gamma function for large $z$
 \begin{equation}
 \Gamma(\nu+z) \sim \sqrt{2\pi}\: e^{(\nu+z-1/2) \ln{z} -z} \quad \mbox{for}
 \quad |z| \rightarrow \infty,
 \end{equation}
 which yields
 \begin{equation}
 \arg\left\{\Gamma(\frac{1}{2}+i\frac{\sqrt{11}}{2}+\alpha)\right\}
  \sim \frac{\sqrt{11}}{2} \ln{\alpha} \quad \mbox{for} \quad \alpha \rightarrow \infty.
 \end{equation}
 Applying this to (\ref{quant})  one gets
 \begin{equation}
 \frac{\alpha_{k+1}}{\alpha_k} \approx e^{\frac{2 \pi}{\sqrt{11}}} \quad \mbox{for} \;\: k \rightarrow \infty.
  \end{equation}
  This formula was useful is providing an initial guess in the the
  numerical root finding procedure for equation (\ref{quant}) for
  large $k$.
\section{Singularities in $D=5$}
Having learned about  self-similar solutions, we are now prepared
to understand the results of numerical studies, first reported
in~\cite{bi_ta}, of  the Cauchy problem for the YM equation in
five space dimensions
\begin{equation}\label{wave5}
w_{tt} = w_{rr} +\frac{2}{r} w_r + \frac{3}{r^2} w (1-w^2) .
\end{equation}
The main goal of these studies was to determine the asymptotics of
blowup. Our numerical simulations were based on finite difference
methods combined with adaptive mesh refinement. The latter were
instrumental in resolving the structure of singularities
developing on vanishingly small scales. We stress that a priori
analytical insight into the problem, in particular the  knowledge
of self-similar solutions was very helpful in interpreting the
numerical results.

We solved equation (\ref{wave5}) for a variety of initial
conditions interpolating between small and large data. A typical
example of such initial data is a Gaussian (ingoing or
time-symmetric)
  of the form \begin{equation}\label{gauss}
 w(0,r)= 1- A r^2 \exp\left[-
\sigma (r-R)^2\right],
\end{equation}
 with adjustable
amplitude $A$ and fixed parameters $\sigma$ and $R$.
 The global behaviour of solutions is qualitatively the same
 for all families of initial data and
  depends critically on the amplitude $A$ (or any other parameter
  which controls the "strength" of initial data).
  For small amplitudes the solutions disperse, that is the energy
  is radiated away to infinity and in any compact region the
  solution approaches the vacuum solution $w=1$. This is in
  agreement with general theorems on global existence for small
  initial data~\cite{kl1}. Heuristically, this follows from the fact
  the for a small amplitude the nonlinearity is dominated by the
  dispersive effect of the linear wave operator.
 For large amplitudes  we observe
the development of two clearly separated regions: an outer region
where the evolution is very slow and a rapidly evolving inner
region where the  solution attains a kink-like shape which shrinks
in a self-similar manner to zero size in a finite time $T$. The
kink is, of course, nothing else but the self-similar solution
$W_0(\frac{r}{T-t})$\footnote{Throughout this section we use the
similarity variable $\eta=\frac{r}{T-t}$ (rather than $x$) and
abuse the notation by writing $W_n(\eta)$ to denote $\tilde
W_n(\eta)=W_n(x)$.} (see Figure~3).
\begin{figure}
\centering
\includegraphics[width=0.8\textwidth]{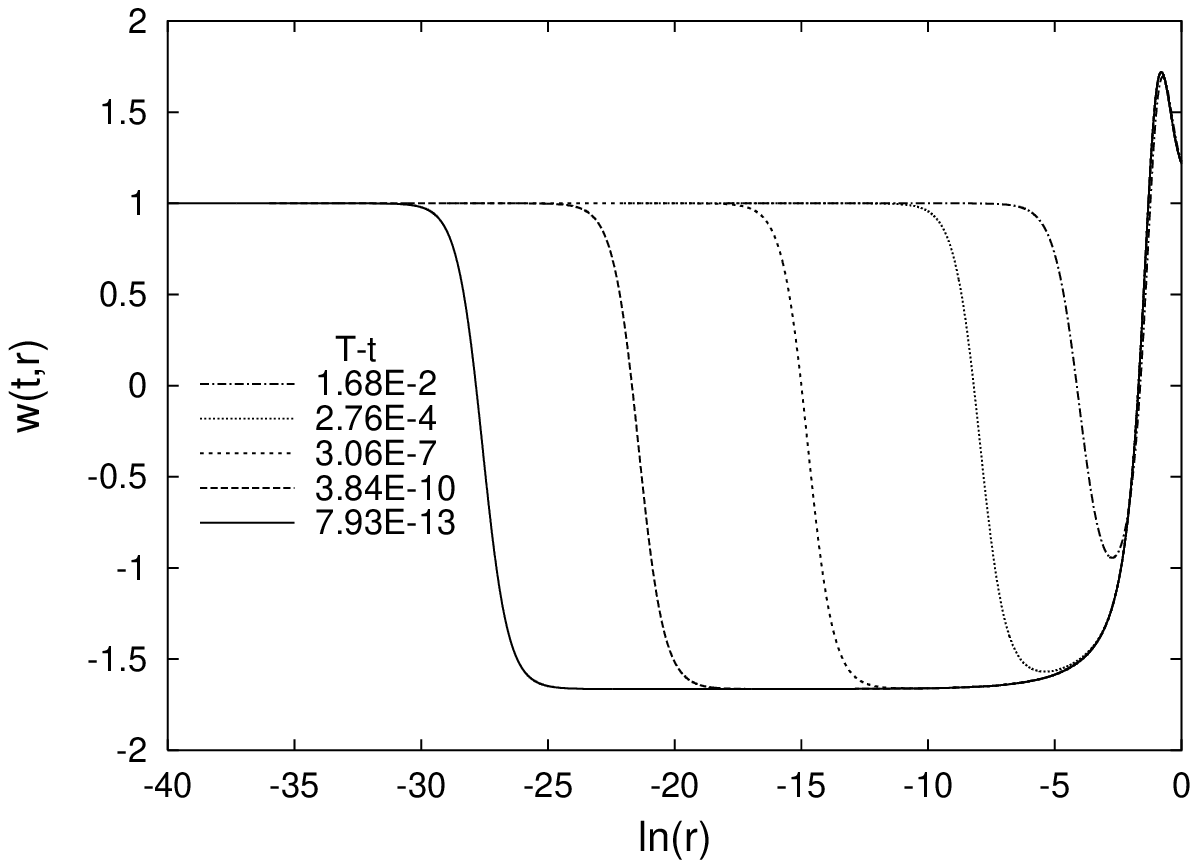}
\includegraphics[width=0.8\textwidth]{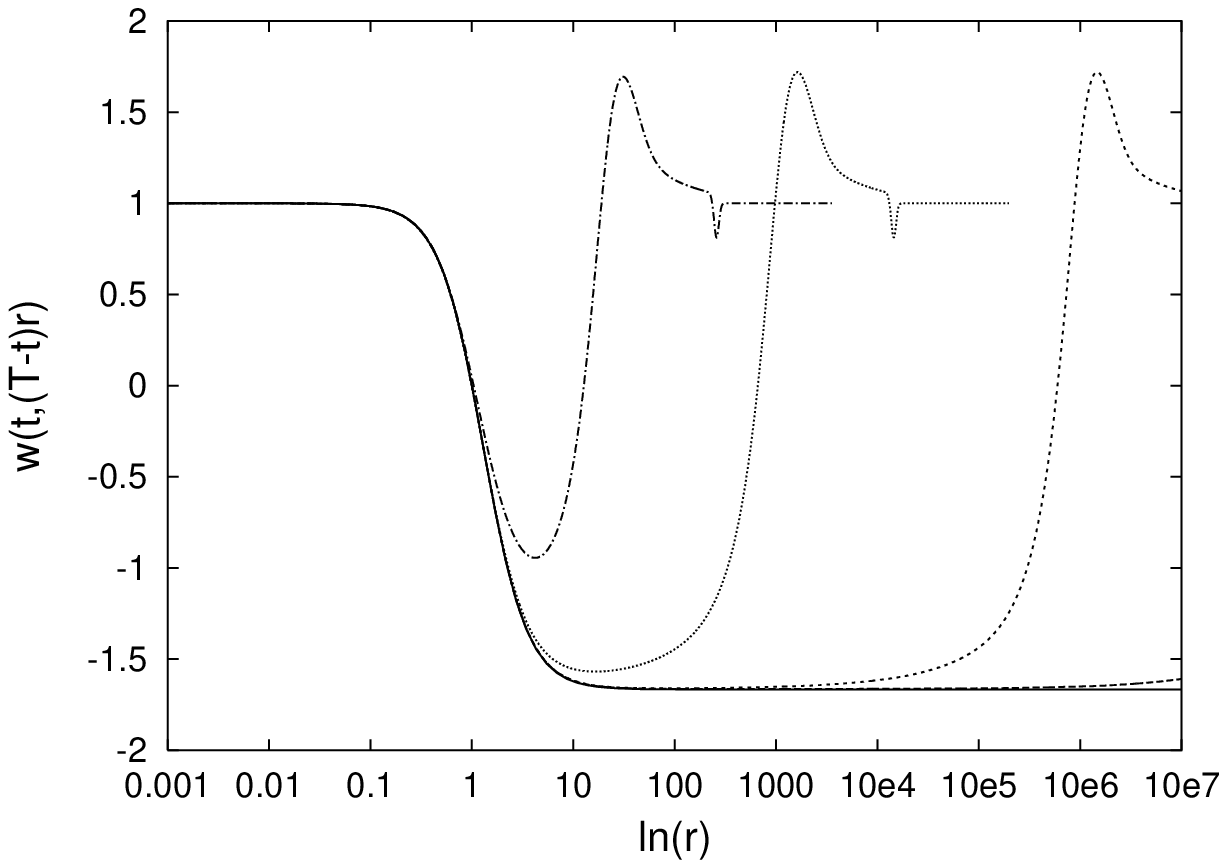}
\caption{\small{The upper plot shows the late time evolution of
time symmetric initial data of the form (\ref{gauss}) with
$\sigma=10, R=2$, and $A=0.2$. As the blowup progresses, the inner
solution gradually attains
 the form of the stable self-similar solution $W_0(r/(T-t))$. The outer solution appears frozen
on this timescale. In the lower plot the rescaled solutions $w(t,
(T-t)) r$ are shown to collapse  to the profile $W_0(r)$ (solid
line).} }\label{fig4}
\end{figure}

We summarize these findings in the following conjecture:
\begin{conj}[On blowup in $D=5$] \emph{Solutions of equation
(\ref{wave5}) corresponding to sufficiently large initial data do
blow up in
 finite time in the
sense that $w_{rr}(t,0)$ diverges as $t \nearrow T$ for some
$T>0$. The universal asymptotic profile of blowup is  given by the
stable self-similar solution:}
\begin{equation}
 \lim_{t\nearrow T} w(t,(T-t) r) = W_0(r).
\end{equation}
\end{conj}
We think that the basic mechanism which is responsible for the
observed asymptotic self-similarity of blowup can be viewed as the
convergence to the lowest "energy" configuration. To see this, let
us rewrite (\ref{wave5})  in terms of the similarity variable
$\eta$ and the slow time $\tau=-\ln(T-t)$ to get
\begin{equation}\label{wave5eta}
w_{\tau\tau} \!+\!w_{\tau} \!+\!2 \eta w_{\eta\tau} \!=\!
(1-\eta^2) \!\left(\!w_{\eta\eta}
\!+\!\frac{2}{\eta}w_{\eta}\!\right)\! +\! \frac{3}{\eta^2} w
(1-w^2).
\end{equation}
In this way  the problem of blowup was converted into the problem
of asymptotic behaviour of solutions for $\tau \rightarrow
\infty$.
 The natural "energy" functional  associated with this problem is
\begin{equation}
K(w)= \int\limits_0^1
 \left (\eta^2 w_{\eta}^2 +\frac{3}{2}\frac{(1-w^2)^2}{1-\eta^2}\right)
 d\eta.
\end{equation}
$K(w)$ has a minimum at the  self-similar solution $W_0$ and
saddle points with $n$ unstable directions at solutions $W_n$ with
$n>0$. Since the wave equation (\ref{wave5eta}) contains a damping
term reflecting an outward flux of energy through the past light
cone of the singularity,  we suspect (but cannot prove) that
$K(w)$ decreases with time. If so, it is natural to expect that
solutions will tend asymptotically to the minimum of $K(w)$.

We already know that solutions with small data  disperse and
solutions with large data blow up. The question is what happens in
between. Using bisection, we found that along each interpolating
family of initial data there is a threshold value of the
parameter, say the amplitude $A^*$, below which the solutions
disperse and above which a singularity is formed. The evolution of
initial data near the threshold was found to go through a
transient phase which is universal, i.e. the same for all
families. This intermediate attractor was identified as the
self-similar solution $W_1$. Having gone through this transient
phase, at the end the solutions leave the intermediate attractor
towards  dispersal or blowup. This behaviour is shown in Figure~4
for the time-symmetric initial data of the form (\ref{gauss}).
\begin{figure}[h]
\centering
\includegraphics[width=0.85\textwidth]{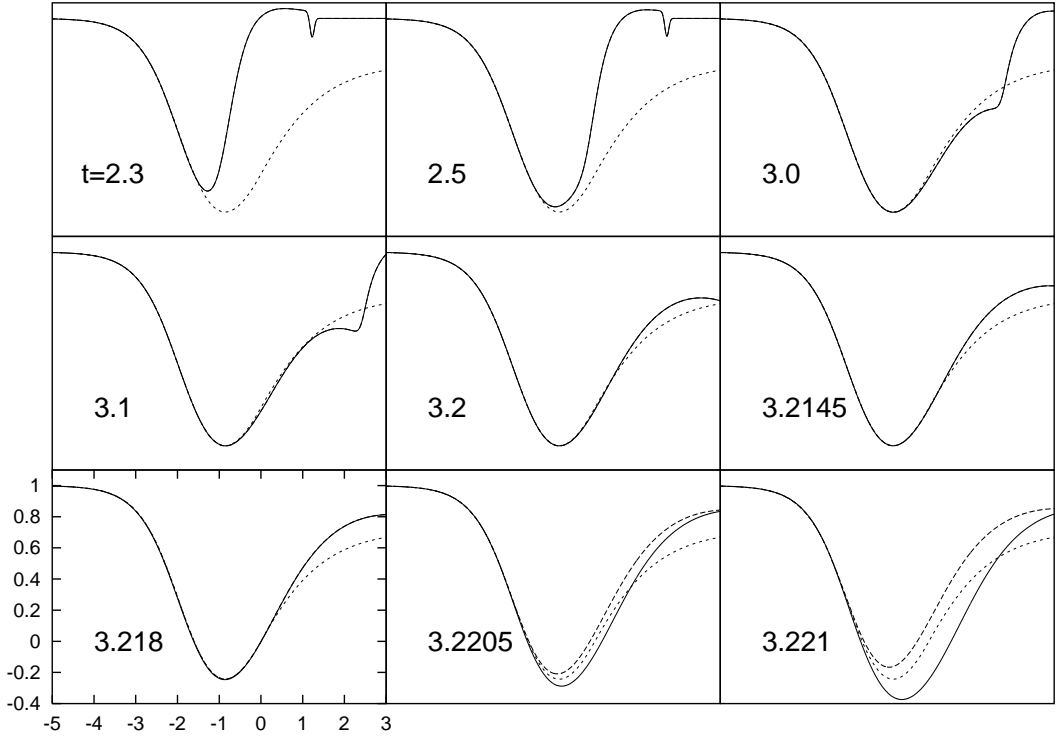}
\vskip 0.15cm
 \caption{\small{The dynamics of time-symmetric initial data of the form (\ref{gauss})
 with amplitudes that are fine-tuned to the threshold of singularity formation. The rescaled solution
$w(t,(T-t)r)$ is plotted against $\ln(r)$ for a sequence~of
intermediate times. Shown (solid and dashed lines) is the pair of
solutions starting with marginally critical
 amplitudes $A=A^*\pm \epsilon$, where $A^*=0.144296087005405$. Since
$\epsilon=10^{-15}$, the two solutions are  indistinguishable on
the first seven frames. The convergence to the self-similar
solution $W_1$ (dotted line) is clearly seen in the intermediate
asymptotics. The last two frames show the solutions departing from
the intermediate attractor towards blowup and dispersal,
respectively.} } \label{fig5}
\end{figure}

 The universality of the  dynamics at the threshold of singularity formation can be
  understood heuristically as follows\footnote{This heuristic
  picture of the dynamics near the threshold,  borrowed from dynamical systems theory, has been first given
  in the context of Einstein's equation -- see Section~6 and~\cite{gu}.}.
  As we showed above, the self-similar solution $W_1$ has exactly one
  unstable mode  -- in other words the
  stable manifold of this solution has codimension one and therefore
  generic one-parameter families of initial data do intersect it. The
  points of intersection correspond to critical initial data that
  converge asymptotically to $W_1$.
  The marginally critical data, by continuity, initially remain close to the stable manifold and approach
$W_1$ for intermediate times but eventually are repelled from its
vicinity  along the one-dimensional unstable manifold (see
Figure~5).
\begin{figure}[h]
\centering
\includegraphics[width=0.8\textwidth]{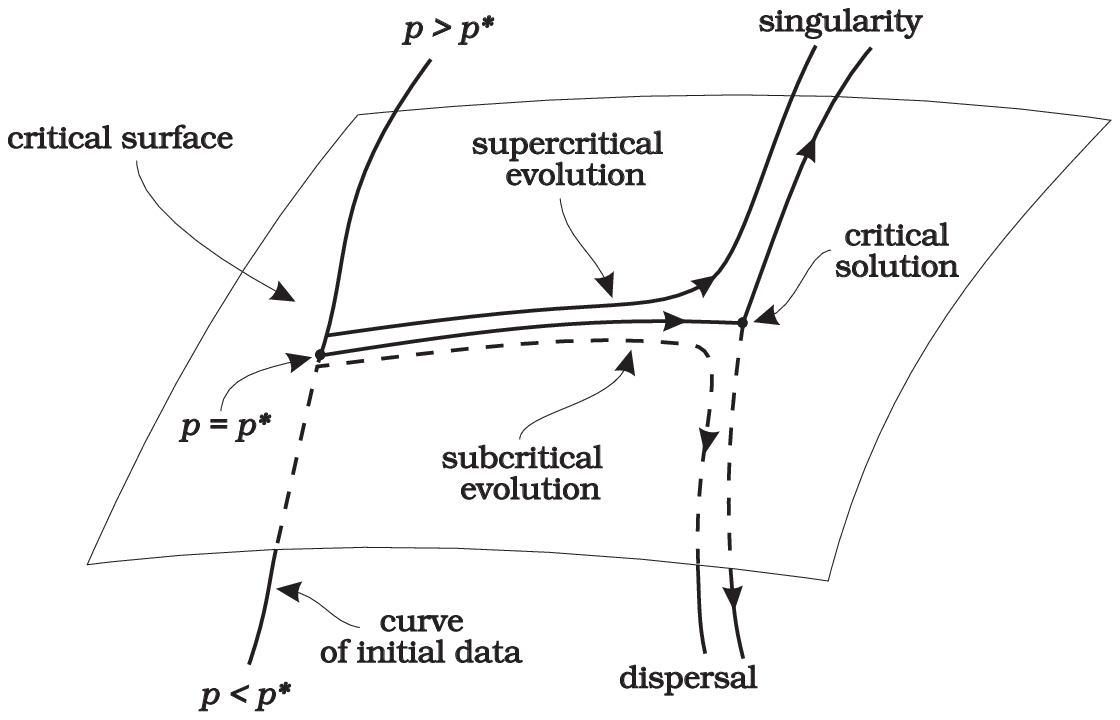}
\caption{\small{A schematic phase space picture of the dynamics at
the threshold of singularity formation.}}
\end{figure}

According to this picture the universality of the nearly critical
dynamics follows immediately from the fact that the same unstable
mode dominates the evolution of all solutions. More precisely, the
evolution of marginally critical  solutions in the
  intermediate asymptotics can be approximated as
  \begin{equation}\label{inter}
  w(t,r) = W_1(\eta) +
   c(A) (T-t)^{-\alpha_1-1} v_(\eta)+ \mbox{radiation} ,
  \end{equation}
  where $v_1$ is the single unstable mode with the eigenvalue
  $\alpha_1=4$.
  The small constant $c(A)$, which is the only vestige
   of the initial data, quantifies an admixture of the unstable mode -- for
   precisely critical data $c(A^*)=0$. The time of departure from
   the intermediate attractor is determined  by the time $t^*$ in which the unstable mode grows to a
  finite size, i.e.,   $c(A) (T-t^*)^{-\alpha_1-1} \sim O(1)$. Using
  $c(A)\approx c'(A^*) (A-A^*)$ and substituting $\alpha_1=4$, we get  $T-t^* \sim
  |A^*-A|^{1/5}$.
 Various scaling laws can be derived
from this. For example, consider solutions with marginally
sub-threshold amplitudes $A=A^*-\epsilon$. For such solutions the
energy density
\begin{equation}\label{density}
e(t,r)= \frac{w_t^2}{r^2}+ \frac{w_r^2}{r^2}+ \frac{3(1-w^2)^2}{2
r^4} \end{equation}
 initially grows at the
center, attains a maximum at a certain time $\approx t^*$ and then
drops to zero. Substituting (\ref{inter}) into (\ref{density}) we
get that $e(t,0)\sim(T-t)^{-4}$, and hence $e(t^*,0) \sim
\epsilon^{-4/5}$.
 \section{Connection with critical phenomena in gravitational
 collapse}
 The behaviour of solutions near the threshold of singularity
 formation described above
shares many features with  critical phenomena at the threshold of
black hole formation in  gravitational collapse. To explain these
similarities, we now briefly recall the phenomenology and
heuristics of the critical gravitational collapse.
Consider  a spherical shell of matter and let it collapse under
its own weight. The dynamics of this process, modelled by
Einstein's equations, can be understood intuitively in terms of
the competition between gravitational attraction and repulsive
internal forces (due, for instance, to kinetic energy of matter or
pressure). If the initial configuration is dilute, then the
repulsive forces "win" and the collapsing matter will rebound or
implode through the center, and eventually will disperse. On the
other hand, if the density of matter is sufficiently large, some
fraction of the initial mass will form a black hole. Critical
gravitational collapse occurs when the attracting and repulsive
forces governing the dynamics of this process are almost in
balance, or in other words, the initial configuration is near the
threshold of black hole formation. The systematic studies of
critical gravitational collapse were launched  in the early
nineties by the seminal paper by Choptuik \cite{ch} in which he
investigated numerically the collapse of a self-gravitating
massless scalar field.

Evolving  initial data fine-tuned to the border  between
no-black-hole and black-hole spacetimes, Choptuik found the
following unforseen phenomena near the threshold:\\ (i)
universality:  all initial data which are near the black hole
threshold go through  a universal transient period in their
evolution during which they approach a certain intermediate
attractor, before eventually dispersing or forming a black hole.
This universal intermediate attractor is usually referred to as
the critical solution.\\ (ii) discrete self-similarity: the
critical solution is discretely self-similar, that is it is
invariant under dilations by a certain fixed factor $\Delta$
called the echoing period.\\
(iii) black-hole mass scaling:  for initial data that do form
black holes, the masses of black holes  satisfy the power law
$M_{bh} \sim \epsilon^{\gamma}$ where $\epsilon$ is the distance
to the threshold and $\gamma$ is a universal (i.e., the same for
all initial data) critical exponent.
 Thus, by fine tuning to the threshold  one can
make an arbitrarily tiny black hole. Put differently, there is no
mass gap at the transition between black-hole and no-black-hole
spacetimes.

What Choptuik found for the scalar field, has been later observed
 in many other models of gravitational collapse,  although the symmetry of the
critical solution itself was found to depend on the model: in some
cases the critical solution  is self-similar (continuously or
discretely), while in other cases the critical solution is static
(or periodic). In the latter case black hole formation turns on
with finite mass. These two kinds of critical behaviour are
referred to as the the type II or type I criticality,
respectively, to emphasize the formal analogy with second and
first order phase transitions in statistical physics. We refer the
interested reader to~\cite{gu} for an excellent review of the
growing literature on critical gravitational collapse.

The present understanding of critical behaviour in gravitational
collapse is based on the same phase space picture as in Figure~5,
that is, it is associated with the existence of a critical
solution with exactly one unstable mode. This picture  leads to
some quantitative predictions. In particular, in the case of type
II critical collapse, an elementary dimensional analysis shows
that the critical exponent $\gamma$ in the power law $M_{bh} \sim
\epsilon^{\gamma}$ is a reciprocal of the unstable eigenvalue of
the critical solution.

By now, the similarities between type II critical gravitational
collapse and the dynamics at the threshold of singularity
formation in the $5+1$ YM equations should be evident. This
analogy, together with similar results for wave maps in $3+1$
dimensions~\cite{bi_ch_ta1},~\cite{li_hi_is}, shows that the basic
properties of critical collapse, such as universality, scaling,
and self-similarity, first observed for Einstein's equations,
actually have nothing to do with gravity and seem to be robust
properties of supercritical nonlinear wave equations. The obvious
advantage of toy models, such as the one presented in this paper,
is their
 simplicity which allowed to get a much
better analytic grip on critical phenomena than in the case of
Einstein's equations; in particular, it was possible to prove
existence  of the critical solution. The only characteristic
property of type II critical collapse which so far has not found
in simpler models (besides, of course, the absence of black holes
which are replaced by singularities) is discrete self-similarity
of the critical solution. It would be very interesting to design a
toy model which exhibits discrete self-similarity at the threshold
for singularity formation because this could give us insight into
the origin of this mysterious symmetry.

\section{Singularities in $D=4$}
In this section we consider the Cauchy problem for the YM equation
in four space dimensions
\begin{equation}\label{wave4}
w_{tt} = w_{rr} +\frac{1}{r} w_r + \frac{2}{r^2} w (1-w^2).
\end{equation}
We begin by recalling some facts concerning equation (\ref{wave4})
which are be important in understanding the dynamics of
singularity formation. First, we note that, in contrast to $D=5$,
there are no smooth self-similar solution in $D=4$.
 This follows from the fact that in $D$
dimensions the local solutions of equation (\ref{sseq}) near the
past light cone behave as $(1-\eta^2)^{\frac{D-3}{2}}$, hence they
are not smooth if $D$ is even (in particular, they are not
differentiable in $D=4$). Although such singular self-similar
solutions do exist, they cannot develop from smooth initial data
and therefore they are not expected to participate in the
dynamics.

Second, $D=4$ is the critical dimension in the sense that the
energy (\ref{energy}) does not change under scaling. This means
that, even though the model is scale invariant, a nontrivial
finite energy static solution may exist\footnote{ Another way of
seeing this is to notice that only in $D=4$ the YM coupling
constant $e^2$ provides the scale of energy.}.
In fact, such a static solution is well known
\begin{equation}\label{instanton}
W_S(r)=\frac{1-r^2}{1+r^2} .
\end{equation}
This is the  instanton in the four-dimensional euclidean YM
theory. Of course, by
 reflection symmetry, $-W_S(r)$ is also the solution. Since the model is scale invariant, the
solution $W_S(r)$
 generates an orbit
of static solutions  $W_S^{\lambda}(r)=W_S(r/\lambda)$, where
$0<\lambda<\infty$.

  To analyze the  linear stability of the instanton, we
   insert
  $w(t,r)=W_S(r)+e^{ikt} v(r)$ into (\ref{wave4}) and linearize. In this way we get the
  eigenvalue problem (the radial Schr\"odinger equation)
\begin{equation}\label{pertstat}
  \left(-\frac{d^2}{dr^2}-\frac{1}{r}\frac{d}{dr} +  V(r)\right) v = k^2
  v,\qquad V(r)=-\frac{2 (1-3 W_S^2)}{r^2}.
\end{equation}
This problem has a zero eigenvalue $k^2=0$ which follows from
scale invariance.  The corresponding eigenfunction (so called zero
mode) is determined by
  the perturbation generated by scaling
\begin{equation}\label{trick}
v_0(r)=-\frac{d}{d \lambda} W_S^{\lambda}(r)
\Bigr\rvert_{\lambda=1} = r W_S'(r) = \frac{4 r^2}{(1+r^2)^2}.
\end{equation}
 Since the zero mode $v_0(r)$ has no
nodes, it follows  by the standard result from Sturm-Liouville
theory that there are no negative eigenvalues, and \emph{eo ipso}
no unstable modes around $W_S(r)$. Thus, the instanton is
marginally stable. Note that the zero eigenvalue lies at the
bottom of the continuous spectrum $k^2 \geq 0$, hence there is no
spectral gap in the problem.

After these preliminaries, we return to the discussion of the
Cauchy problem for equation (\ref{wave4}). For small energies the
solutions disperse, in agreement with general theorems. For large
energies, at first sight the global behaviour seems similar to the
$D=5$ case -- as before, near the center the solution attains the
form of a kink which shrinks to zero size. However, this
similarity is superficial because now the kink is not a
self-similar solution (as no such solution exists). It turns out
(see Figure~6) that the kink has the form of
 the \emph{scale-evolving} instanton
\begin{equation}\label{adiabatic}
 w(t,r) \approx W_S\left(\frac{r}{\lambda(t)}\right),
\end{equation}
 where a scaling factor $\lambda(t)$ is a positive function which
tends to zero as $t\rightarrow T$.
\begin{figure}
\centering
\includegraphics[width=0.8\textwidth]{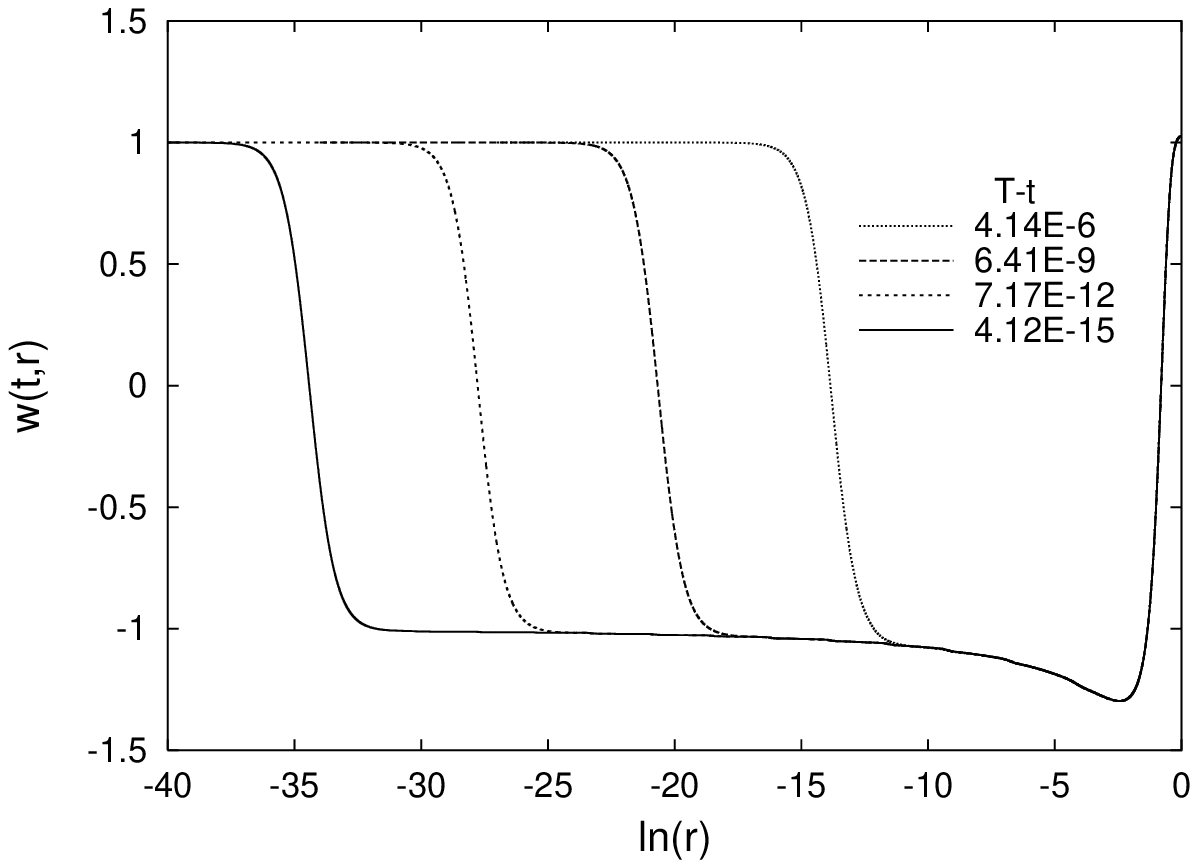}
\includegraphics[width=0.8\textwidth]{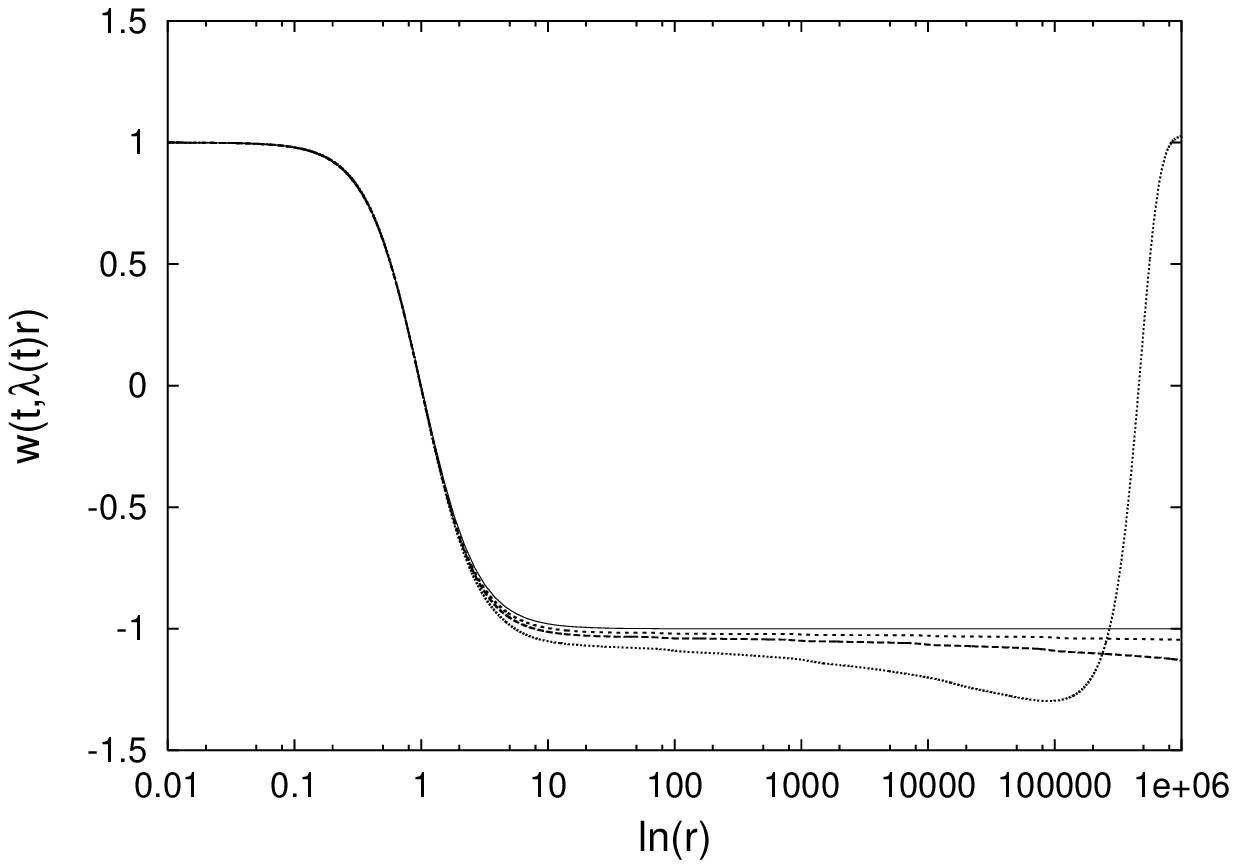}
\caption{\small{The upper plot shows the formation of a
singularity for large initial data of the form (\ref{gauss}) with
$A=0.5$. The inner solution has the form of the scale-evolving
instanton $W_S(r/\lambda(t))$ with the scale factor $\lambda(t)$
going to zero slightly faster than linearly. In the lower plot the
rescaled solutions are shown to collapse to the profile of the
instanton $W_S(r)$ (solid line). }}
\end{figure}
 We summarize these findings in
the following conjecture: \begin{conj}[On blowup in $D=4$]
 \emph{Solutions of equation (\ref{wave4}) with sufficiently large energy
 do blow up in finite time in the sense
that $w_{rr}(t,0)$ diverges as $t \nearrow T$ for some $T>0$. The
universal asymptotic profile of blowup is given by the instanton.
More precisely, there exists a positive function
$\lambda(t)\searrow 0$ for $t\nearrow T$ such that}
\begin{equation}\label{conj2}
 \lim_{t\nearrow T} u(t,\lambda(t) r) = W_S(r).
\end{equation}
\end{conj}
The key question which is left open in this conjecture is: what
determines the evolution of the scaling factor $\lambda(t)$; in
particular, what is the asymptotic behaviour of $\lambda(t)$ for
$t\rightarrow T$?  Numerical evidence shown in Figure~7 suggests
that the rate of blowup goes asymptotically to zero, that is
($\;\dot{}=d/dt$)
\begin{equation}\label{rate}
 \lim_{t\rightarrow T} \frac{\lambda(t)}{T-t} = -\lim_{t\rightarrow T} \dot \lambda
 =0,
\end{equation}
but it seems very hard to determine an exact  asymptotics of
$\lambda(t)$ from pure numerics\footnote{This issue is also
discussed by Linhart and Sadun in~\cite{li_sa}.}.

\begin{figure}[h]
\centering
\includegraphics[width=0.8\textwidth]{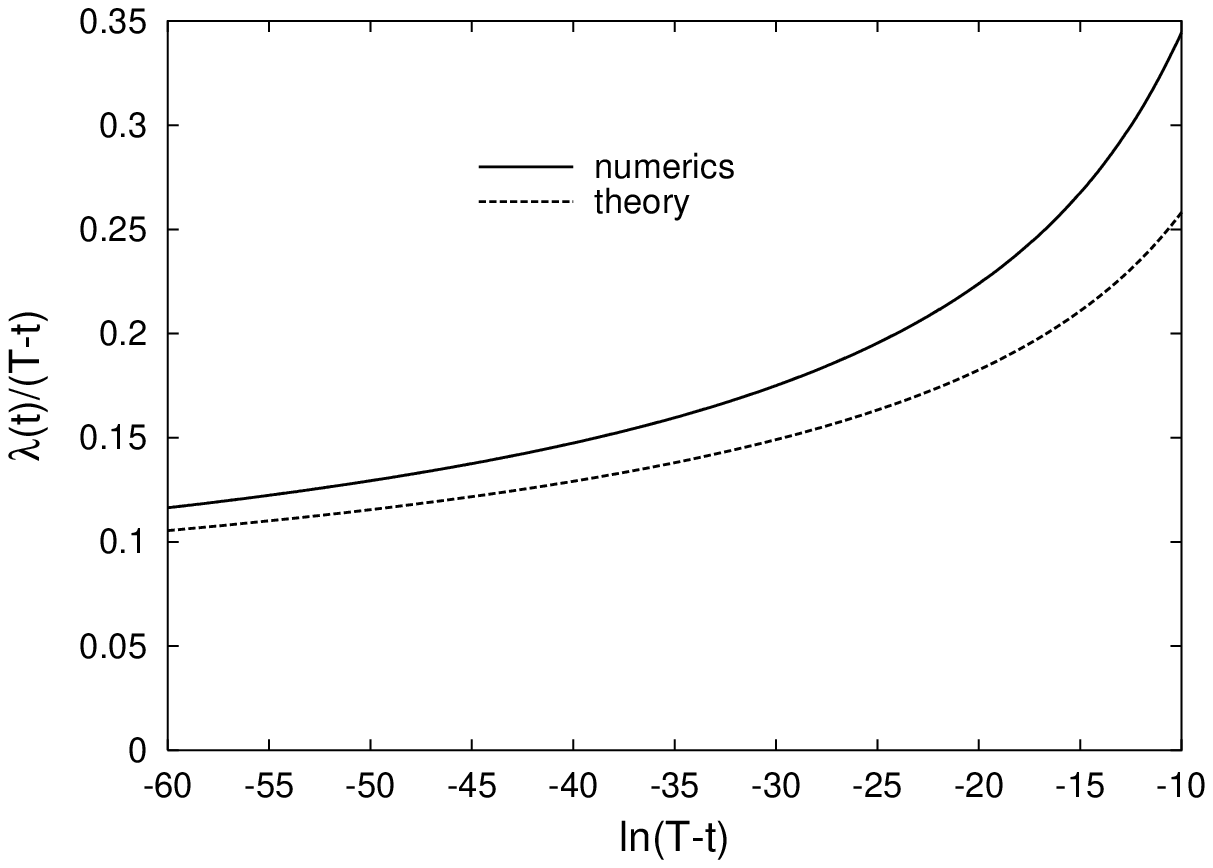}
\caption{\small{Comparison of the numerically  computated scaling
factor divided by $T-t$ (for the same data as in Figure~6) with
the analytic formula $\frac{\lambda(t)}{T-t}=\sqrt{\frac{2}{3}}
(-\ln(T-t))^{-1/2}$. }}
\end{figure}

Recently, an analytical approach to this problem  has been
suggested in~\cite{bi_ov_si}. Below we sketch the main idea of
this approach.
 Let
$\mathcal{M}\equiv \{W_S(r/\lambda)| \lambda\in \mathbb{R^+}\}$ be
a manifold of rescaled instantons (a one-dimensional center
manifold). Assuming that a solution is in a neighbourhood of
$\mathcal{M}$, we decompose it as
\begin{equation}\label{decomp}
w=W_S(\eta) + v(t,\eta), \qquad \eta=r/\lambda(t).
\end{equation}
Here $v$ represents a small deviation of the solution from
$\mathcal{M}$ and $\lambda$ is the collective coordinate on
$\mathcal{M}$. To fix the splitting between these two parts we
require that $P v=0$, where $P$ is the projection on
$\mathcal{M}$.  Plugging (\ref{decomp}) into (\ref{wave4}) we get
($'=d/d\eta$)
 \begin{equation}\label{main2}
\lambda^2
\ddot v - 2 \eta \dot \lambda \lambda \dot v' + L v + N(v) =
\lambda \ddot \lambda \eta W_S' -\dot \lambda^2 (\eta^2  W_S '' +
2 \eta W_S'),
\end{equation} where $L$ is the linear
perturbation operator about the instanton
\begin{equation}\label{L}
 L  = -\frac{\partial^2}{\partial\eta^2}-\frac{1}{\eta}\frac{\partial}{\partial\eta} - \frac{2 (1-3 W_S^2)}{\eta^2},
\end{equation} and
\begin{equation}\label{N}
 N(v)=\frac{6 W_S}{\eta^2} v^2 +
\frac{2}{\eta^2} v^3.
\end{equation}
It is clear from (\ref{main2}) that $v=O(\dot\lambda^2)$, hence
for $v$ to decay to zero as $t\rightarrow T$, the rate of blowup
must go to zero as well. We stress this point to emphasize that
the linear evolution of $\lambda(t)$, predicted for example by the
geodesic approximation, is inconsistent with Conjecture~2.
 Next, by projecting equation (\ref{main2})
on $\mathcal{M}$ and in the orthogonal direction, we get a coupled
system consisting of a nonhomogeneous wave equation for $v$ and an
ordinary differential equation for $\lambda$. Solving the first
equation for $v$ and plugging the result into the second equation,
we obtain in the lowest order the following modulation equation
\begin{equation}\label{effective}
\lambda \ddot \lambda = \frac{3}{4} \dot \lambda^4.
\end{equation}
From this we get the leading order asymptotics for $t \rightarrow
T$
 \begin{equation}\label{sigal}
 \lambda(t) \sim \sqrt{\frac{2}{3}}
\frac{T-t}{\sqrt{-\ln(T-t)}}.
\end{equation}
As shown in Figure~7 this result is in rough agreement with
numerics. There are many possible sources of the apparent
discrepancy. On the numerical side there are discretization
errors, an error in estimating the blowup time, or errors in
computing $\lambda$ from the data. On the analytical side, there
might be corrections to (\ref{sigal}) coming from the bounded
region expansion and, more importantly, from the far field
behaviour\footnote{The derivation of (\ref{effective}) is not
quite straightforward because of the presence of infrared
divergencies which need to be regularized.}. Finally, and in our
opinion most likely reason of discrepancy is that the solution
shown in Figure~7 has not yet reached
 the truly asymptotic regime and consequently the higher order
corrections to formula (\ref{sigal}) are still significant.

The issue of blowup rate is closely related to the problem
 of energy concentration in the singularity. To explain this,
   we define  the kinetic and the potential energies at time $t<T$ inside the past light cone
of the singularity
\begin{equation}\label{energy_cone}
E_K(t) = 6\pi^2 \!\int\limits_0^{T-t} w_t^2 r dr, \qquad E_P(t) =
6\pi^2\! \int\limits_0^{T-t} \left(w_r^2 +
\frac{(1-w^2)^2}{r^2}\right) r dr.
\end{equation}
Substituting (\ref{adiabatic}) into (\ref{energy_cone}) we obtain
\begin{equation}\label{en_conc}
E_K(t) = 6\pi^2 \!\dot\lambda^2
\int\limits_0^{\frac{T-t}{\lambda(t)}} {W'_S}^2 r dr, \qquad
E_P(t) = 6\pi^2 \int\limits_0^{\frac{T-t}{\lambda(t)}}
\left({W'_S}^2 + \frac{(1-W_S^2)^2}{r^2}\right) r dr.
\end{equation}
Assuming (\ref{rate}), this implies that
\begin{equation}
\lim_{t \rightarrow T} E_K(t)= 0, \qquad \lim_{t \rightarrow T}
E_P(t) = 16 \pi^2.
\end{equation}
Thus, the energy equal to the energy of the instanton gets
concentrated in the singularity.  This means that in the process
of blowup the excess energy must be radiated away from the inner
region as the solution converges to the instanton.

It is worth pointing out that the concentration of energy is a
necessary condition for blowup in the critical dimension. To see
this, suppose that the solution blows up at time T and  assume for
contradiction that $\lim_{t \rightarrow T} E(t)= 0$. Then, by
choosing a sufficiently small  $\epsilon>0$ we can have
$E(T-\epsilon)$ arbitrarily small. This implies, by causality and
global existence for small energy data, that the solution exists
globally in time, contradicting the assumption. For radial
equations of the type
\begin{equation}\label{struwe}
w_{tt} = w_{rr} +\frac{1}{r} w_r + \frac{f(w)}{r^2},
\end{equation}
a stronger result was proved by Struwe~\cite{st} who basically
showed that if the solution of (\ref{struwe})  blows up, then it
must do so in the manner described in Conjecture~2. In this sense
formation of singularities is intimately tied with the existence
of a static solution.

Finally, we address briefly the issue of the threshold of
singularity formation. Using the technique described in Section~5,
along each interpolating family of initial data we can determine
the critical point separating blowup from dispersal. However, in
contrast to the $D=5$ case, we see no evidence for the existence
of an intermediate attractor in the evolution of nearly critical
data. This fact, together with a similar result for $2+1$
dimensional wave maps~\cite{bi_ch_ta2} suggests that in the
critical dimension the transition between blowup and dispersal is
not governed by any critical solution. We suspect that the
evolution of precisely critical initial data still has the  form
(\ref{adiabatic}) but the dynamics of the scaling factor is
different than in (\ref{sigal}). This belief is based on the fact
that in the evolution of marginally critical initial data we can
clearly distinguish a transient phase during which $\lambda(t)$
drops very quickly and then, either reaches a minimum and starts
growing (in the case of dispersal), or keeps decreasing with the
"normal" rate (in the case of blowup).

\section{Conclusions}
There are two main lessons that we wanted to convey in this
survey. The first lesson is that there are striking analogies
between major evolution equations. In particular, the mechanism of
blowup  to a large extent is determined by the criticality class
of the model. These analogies can be used to get insight into hard
problems (such as singularity formation for Einstein's equations)
by studying toy models which belong to the \emph{same} criticality
class. This approach is in the spirit of  general philosophy
expressed by David Hilbert in his famous lecture delivered before
the International Congress of Mathematicians at Paris in
1900~\cite{hi}: \emph{"In dealing with mathematical problems,
specialization plays, as I believe, a still more important part
than generalization. Perhaps in most cases where we seek in vain
the answer to a question, the cause of the failure lies in the
fact that problems simpler and easier than the one in hand have
been either not at all or incompletely solved. All depends, then,
on finding out these easier problems, and on solving them by means
of devices as perfect as possible and of concepts capable of
generalization."}

The second lesson is concerned with the interplay between
numerical and analytical techniques.
 Accurate and reliable
numerical simulation of singular behaviour is difficult and hard
to assess. In order to keep track of a singularity developing on
exceedingly small spatio-temporal scales, one needs sophisticated
techniques such as adaptive mesh refinement. For these techniques
the convergence and error analysis are lacking so extreme care is
needed to make sure that the computed singularities are not
numerical artifacts. For this reason, in order to feel confident
about numerics it is important to have some analytical
information, like existence of self-similar solutions. Without a
theory, simulations alone do not provide ample evidence for the
existence of a singularity. We believe that the interaction
between numerical and analytical techniques, illustrated here by
the studies of blowup, will become more and more important in
future as we begin to attack more difficult problems.

\vskip 0.2cm  \emph{Acknowledgment.} I am grateful to Michael
Sigal and Yu. N. Ovchinnikov for permission to announce here the
results of yet unpublished joint work. I thank Zbislaw Tabor for
providing me with data for Figure~7. I acknowledge the hospitality
of the Albert Einstein Institute for Gravitational Physics in Golm
and the Erwin Schr\"odinger Institute for Mathematical Physics in
Vienna, where parts of this paper were produced.

\end{document}